\documentclass[letterpaper, 10 pt, conference]{ieeeconf}  % Comment this line out if you need a4paper

\IEEEoverridecommandlockouts                              % This command is only needed if 
\usepackage{amsmath} % assumes amsmath package installed
\usepackage{xcolor}
\usepackage{graphics} % for pdf, bitmapped graphics files
\usepackage{epsfig} % for postscript graphics files
\usepackage{amssymb}  % assumes amsmath package installed
\usepackage{enumerate}
\usepackage{caption}
\usepackage{subcaption}
 \newtheorem{theorem}{Theorem}[section]
 \newtheorem{proposition}[theorem]{Proposition}
 \newtheorem{corollary}[theorem]{Corollary}
\newtheorem{lemma}[theorem]{Lemma}

\newtheorem{definition}{Definition}
 \newtheorem{remark}[theorem]{Remark}
\usepackage[ruled,linesnumbered]{algorithm2e}
\usepackage{hyperref}
\hypersetup{pdfborder={0 0 0},
            breaklinks=true,%
            bookmarksnumbered=true,%
            colorlinks=true,%
            citecolor=blue,%
            linkcolor=blue,%
}

\def\qed{\hfill\ensuremath{\square}}

\title{\LARGE \bf
Perimeter Defense using a Turret with Finite Range and Service Times
}

\author{Shivam Bajaj$^1$, Shaunak D. Bopardikar$^1$, Alexander Von Moll$^3$, Eric Torng$^2$ and David W. Casbeer$^3$% <-this % stops a space
\thanks{$^{1}$S. Bajaj and S. D. Bopardikar are with the Department of Electrical and Computer Engineering, Michigan State University.
}
\thanks{$^{2}$E. Torng is with the Department of Computer Science and Engineering, Michigan State University.}
\thanks{$^{3}$A. Von Moll and D. W. Casbeer are with Control Science
Center, Air Force Research Laboratory.}
\thanks{This research was supported by the Aerospace Systems Technology Research and Assessment (ASTRA) Aerospace Technology Development and Testing (ATDT) program at AFRL under contract number FA865021D2602. Approved for public release: distribution unlimited, case number: AFRL-2023-0444.}
}

\begin{document}

\maketitle
\thispagestyle{empty}
\pagestyle{empty}

%%%%%%%%%%%%%%%%%%%%%%%%%%%%%%%%%%%%%%%%%%%%%%%%%%%%%%%%%%%%%%%%%%%%%%%%%%%%%%%%
\begin{abstract}
We consider a perimeter defense problem in a planar conical environment comprising a single turret that has a finite range and non-zero service time. The turret seeks to defend a concentric perimeter against $N\geq 2$ intruders. Upon release, each intruder moves radially towards the perimeter with a fixed speed. To capture an intruder, the turret’s angle must be aligned with that of the intruder’s angle and must spend a specified service time at that orientation. We address offline and online versions of this optimization problem. Specifically, in the offline version, we establish that in general parameter regimes, this problem is equivalent to solving a Travelling Repairperson Problem with Time Windows (TRP-TW). We then identify specific parameter regimes in which there is a polynomial time algorithm that maximizes the number of intruders captured. In the online version, we present a competitive analysis technique in which we establish a fundamental guarantee on the existence of at best $(N-1)$-competitive algorithms. We also design two online algorithms that are provably $1$ and $2$-competitive in specific parameter regimes.
\end{abstract}

%%%%%%%%%%%%%%%%%%%%%%%%%%%%%%%%%%%%%%%%%%%%%%%%%%%%%%%%%%%%%%%%%%%%%%%%%%%%%%%%

\section{Introduction}
This work considers an offline as well as an online version of a perimeter defense problem in a planar conical environment. The environment consists of a concentric perimeter which is guarded by a turret located at the origin. Intruders move radially inwards with fixed speed and seek to breach the perimeter. The turret, having a finite range and service time, can turn with bounded angular speed and seeks to capture the intruders before they reach the perimeter.  

Perimeter defense problems have recently gained a lot of attention. After the seminal work in \cite{isaacs1999differential}, these problems have been mostly formulated as a pursuit evasion differential game commonly studied as reach avoid games \cite{selvakumar2019feedback}. A typical approach requires
computing solutions to the Hamilton-Jacobi-Bellman-Isaacs
equation, which is tractable only in low dimensional state spaces \cite{fisac2015reach}. A particular class of perimeter defense problems has the defenders constrained to be on the perimeter \cite{shishika2018local-game,shishika2019perimeter-defense}. We refer to  \cite{shishika2020review} for a review of such perimeter defense games. Recently, \cite{akilan2017zero-sum} introduced a differential game between a turret and a mobile intruder with an instantaneous cost based on the angular separation between the two. A similar problem setup with the possibility of retreat was considered in
% Optimality conditions, policies and three singular surfaces were determined, out of which two were analyzed and the third was computed numerically in  \cite{vonmoll2020attacker}. 
% This work was then extended to consider a possibility of retreat for the attacker in
\cite{vonmoll2020optimal,vonmoll2021turret}. Further, \cite{vonmoll2022circular} and \cite{vonmoll2022turret-runner-penetrator} considered a scenario in which the turret seeks to align its angle to that of the intruders in order to neutralize an attacker. Other recent works include \cite{guerrerobonilla2021robust} and \cite{lee2021optimal} which consider an approach based on control barrier function or a convex shaped perimeter, respectively. All of these works assume availability of some information, such as locations or total number, about the intruders a priori.

Dynamic vehicle routing problems (DVR) is a class of online optimization problems that require the route of the vehicle to be re-planned as information is revealed gradually over time \cite{psaraftis1988dynamic, bertsimas1991stochastic}. The most relevant works in this area are TRP-TW problems in which most of the works consider either zero or stochastic service times \cite{miranda2016vehicle,gao2020approximation,pavone2009stochastic,bar2005TSPTW,gutierrez2006whack}.  Generally, the aim in such problems is to find a route through a static input in order to minimize or maximize the cost. Conversely, in perimeter defense scenarios, the input (intruders) moves toward a specified region (perimeter) and hence, this problem is more challenging than the former. With the assumption that the arrival process of the intruders is stochastic, \cite{bajaj2019dynamic,smith2009dynamic,macharet2020adaptive} consider the perimeter defense problem as a vehicle routing problem and provide insights into the average case analysis of such problems. An important distinction between the online setup of this work from \cite{OLDTRP_TW_Azar_2017} is that the time taken by the intruders to reach the perimeter is fixed and not the part of the input.

Although these works provide valuable insights, they either do not scale well with an arbitrary number of intruders released online or do not account for scenarios in which intruders may coordinate their arrival to overcome the defense.

% Most prior work has either focused on determining an optimal strategy for few intruders or intruders generated by a stochastic process. These approaches either do not scale well with an arbitrary number of intruders released online or do not account for the worst-case in which intruders may coordinate their arrival to overcome the defense.

This work considers a perimeter defense problem with a single turret and $N$ intruders. 
We consider both offline and online setups.
For the offline setup, we characterize the complexity of the problem and provide a polynomial time algorithm, in a specified parameter regime, that maximizes the number of intruders captured out of the total $N$ intruders. 
% \ekt{[This makes it sound like we can always capture all $N$ intruders; I think we have to qualify this statement.]}
For the online setup, we design online algorithms and provide analytical bounds on their performance in the worst-case.
We adopt a \emph{competitive analysis} perspective to evaluate the performance of online algorithms in the worst-case \cite{sleator1985amortized}. Under this paradigm, an online algorithm $\mathcal{A}$'s performance is measured using the notion of \emph{competitive ratio}: the ratio of the optimal (possibly non-causal) algorithm’s performance and algorithm $\mathcal{A}$’s performance for a worst-case input sequence for algorithm $\mathcal{A}$.
An algorithm is $c$-competitive if its competitive ratio is no larger than $c$, i.e., its performance is guaranteed to be within a factor $c$ of the optimal.

Previously, we introduced the perimeter defense problem for a single mobile defender in linear environments using competitive analysis \cite{bajaj2021linear}, which was later followed by~\cite{bajaj2022conical} for conical environments. The key distinction of this work from our previous work is that we consider a different model for the defender, i.e., a turret. Additionally, we also consider an offline setup which was not considered previously.

Our general contribution is that we analyze an offline as well as online perimeter defense problem using a turret that has a finite range and non-zero service time in a planar conical environment of unit radius and angle $2\theta$. The turret seeks to defend a coaxial and concentric perimeter of radius $\rho<1$ by capturing as many intruders as possible. 
For the offline setup, we consider $N$ intruders in the environment at arbitrary given locations. For the online setup, we consider that at most $N$ intruders can be released over time. In the online version, the locations of the intruders and the arrival times are not known to the turret until the intruders are released. In both setups, the intruders move radially towards the perimeter with a fixed speed $v>0$. Our main contributions are as follows. For the offline setup, we establish that the problem is equivalent to solving a TRP-TW. Then, we determine a specific parameter regime which admit a polynomial time optimal solution to the problem. For the online setup, we first characterize a parameter regime in which no algorithm can have a competitive ratio better than $(N-1)$. Next, we design and analyze two classes of online algorithms that are provably $1$ and $2$-competitive in specific parameter regimes.

This paper is organized as follows. In Section \ref{sec:Problem}, we formally define our problem and competitive ratio. Section \ref{sec:P1} provides the analysis on the offline setup and Section \ref{sec:P2} provides the analysis for the online version of the problem. Finally, in Section \ref{sec:conc} we summarize this work and outline directions for future work.

\section{Problem Formulation}\label{sec:Problem}
In this section, we first describe the model and then formally define the offline as well as the online problem considered in this work.
\begin{figure}[t]
    \centering
    \includegraphics[scale=0.2]{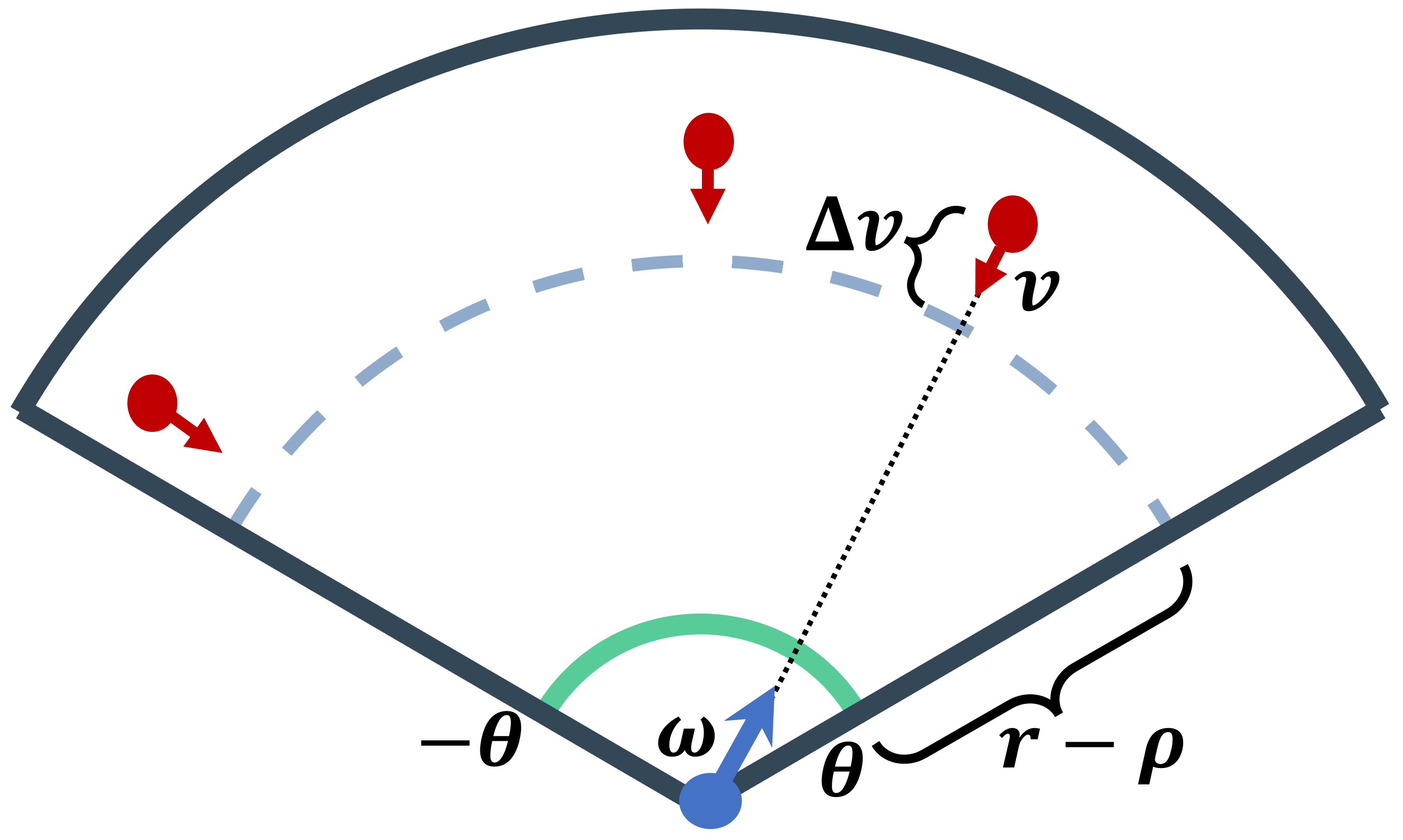}
    \caption{\small Problem Description. The green curve denotes the perimeter and the blue arrow denotes the turret. The blue dashed curve denotes the range of the turret and the red dots denote the intruders. The black dashed line denotes that the turret has locked onto an intruder and the intruder will be captured at $r$ distance away from the origin.}
    \label{fig:prob_desc}
\vspace{-0.2in}
\end{figure}
\subsection{Model}
We consider a planar conical environment (cf. Fig. \ref{fig:prob_desc}) described by $\mathcal{E}(\theta)=\{(y,\alpha) \, : \, 0 < y \leq 1, -\theta\leq\alpha\leq \theta \}$, containing a concentric and coaxial region $\mathcal{R}(\rho,\theta)=\{(z,\alpha) \, : \, 0 < z \leq \rho<1, -\theta\leq\alpha\leq \theta \}$, where angle $\theta$ is measured from the $y$-axis. Note that $(y,\alpha)$ and $(z,\alpha)$ represents locations in polar coordinates. An arbitrary number of intruders are released at the circumference of the environment, i.e., $y=1$ at arbitrary time instants. Upon release, each intruder moves radially  with a fixed speed $v$ toward the perimeter $\partial \mathcal{R}(\theta) = \{(\rho,\alpha) \, : \, -\theta\leq\alpha\leq \theta \}$. Specifically, if the $j$th intruder is released at time $t_j$, then its location is represented by a constant angle $\theta_j$ and its distance $z_j^t$ from the origin satisfying $z_j^t = 1-v(t-t_j), \forall t \in [t_j, t_j+(1-\rho)/v]$. We assume that there are a total of $N>1$ intruders where $N\in \mathbb{N}$. The conical region $\mathcal{R}(\rho,\theta)$ consists of a single turret located at the origin of $\mathcal{E}(\theta)$. Although the turret has information of all intruders that arrive in the environment, it can only neutralize an intruder within its range $r$. Further, the turret requires a fixed \emph{spool up} or service time $\Delta$ to neutralize an intruder. We assume that once the turret spools up, it requires no additional time to neutralize an intruder. This means that the turret may spool up for an intruder if the intruder is located at most $r+\Delta v$ distance away from the origin such that it is captured as soon as it is $r$ distance away from the origin (cf. Fig. \ref{fig:prob_desc}). Thus, the turret is characterized by the following parameters: 
\begin{itemize}
    \item Heading angle ($\gamma_t \in [-\theta, \theta]$): The heading angle defines the direction in which the turret points at time $t$.
    
    \item Angular speed ($\omega$): This is the angular speed with which the turret can turn in either direction. We assume simple kinematics, i.e., $\dot{\gamma}_t = u$, where $u$ is a measurable signal taking values in $[-\omega, +\omega]$.
    
    \item Range ($r \in [\rho, 1]$): This is the radial distance until which the turret can neutralize an intruder.

    \item Service time ($\Delta>0$): The service time corresponds to the spool up time or startup time required by the turret to neutralize an intruder. During the service time the turret's heading angle must not change. Once the spool up time is complete the turret neutralizes an intruder if it is within its range. Note that there is no benefit for the turret to spool up for an intruder located at $(z_t^j,\theta_j)$ if $z_t^j>r+\Delta v$. However, the turret may wait at angle $\theta_j$ until $z_t^j=r+\Delta v$ and then spool up. Finally, the analysis can easily be applied to scenarios in which the turret's service time corresponds to capturing an intruder. In such scenarios, the turret begins capturing an intruder when it is at most $r$ distance away from the origin and finishes capturing once it is $r-\Delta v$ distance away from the origin.
\end{itemize}
The information set $I_t$ available to a turret at time $t$ consists of the locations and release times of each intruder that has been released until time $t$. Using $I_t$, we define a (feedback) control policy for the turret $u(t,\gamma_t,I_t) \, : \, [0,+\infty)\times [-\theta,\theta] \times \Phi_N \to [-\omega, +\omega]$, where $\Phi_N$ represents the set of all possible locations and release times of $N$ intruders until time $t$.  

We now proceed to formally define \emph{capture} of an intruder. 
The turret is said to \emph{lock on} the $j$th intruder, located at $(z_t^j,\theta_j),$ if $\gamma_t=\theta_j$ and the turret decides to capture the intruder, and thus begins spooling up. We assume that once the turret initiates spooling up, it must spend $\Delta$ time units at the same angle. In other words, if the turret locks on to the $j$th intruder at time $t$, then the $j$th intruder is captured and removed from $\mathcal{E}(\theta)$ at time $t+\Delta$. Finally, if there are $1<n\leq N$ intruders collocated at $(z_t^n,\theta_n)$, where $z_t^n\leq r+\Delta v,$ and $\theta_n=\gamma_t$ holds, then the turret requires $n\Delta$ time to capture all intruders. 
% \ekt{[Do we need to specify that the intruders are within range during this time and do not breach?]} 

\subsection{Problem statements}
A \emph{problem instance} $\mathcal{P}$ is characterized by seven parameters that are $\theta,\rho,v,\omega,r,N$ and $\Delta$, where the perimeter's size is normalized by the size of the environment. We consider the following two problems. The first is an offline version which assumes that the $N$ intruders are already present in $\mathcal{E}(\theta)$, i.e., at locations $I_0 := \{(z^0_1, \theta_1), \dots, (z^0_N, \theta_N)\}$. The problem is formally defined as follows:

\textbf{Problem Statement I (P1):} \textit{Given $N$ intruders in an environment $\mathcal{E}(\theta)$ and the initial turret heading of $\gamma_0$, determine a control policy $u^*(t,\gamma_0, \mathcal{I}_0)$ for the turret that maximizes the number of intruders captured.}

The second is an online version in which at most $N$ intruders arrive at arbitrary time instants and locations. We start with some definitions.  

An input sequence $\mathcal{I}$ is a set of 3-tuples comprising: (i) an arbitrary time instant $t\leq T$, where $T$ denotes the final time instant, (ii) the number of intruders $n(t)$ that are released at time instant $t$, and (iii) the release location (radius and angle) of each of the $n(t)$ intruders. Formally,
$\mathcal{I}=\{t,n(t),\{ (1,\alpha_1),(1,\alpha_2),\dots,(1,\alpha_{n(t)}) \}\}_{t=0}^T$, for any $\alpha_l \in [-\theta,\theta]$, where $1\leq l\leq n(t)$. 

An \emph{online algorithm} $\mathcal{A}$ assigns angular velocity with magnitude $\omega$ to the turret at time $t$ as a function of the input instance, or equivalently information set, $I_t\subset \mathcal{I}$ revealed until time $t$. 
An \emph{optimal offline algorithm} is a non-causal algorithm which has complete information of the entire input sequence $\mathcal{I}$ to assign angular velocity to the turret at any time $t$.
% The performance of an online algorithm (resp. optimal offline algorithm) for a problem instance $\mathcal{P}$ is the total number of intruders captured by the turret on an input sequence $\mathcal{I}$. 
Let $m_{\mathcal{A}}(\mathcal{I},\mathcal{P})$ (resp. $m_{\mathcal{O}}(\mathcal{I},\mathcal{P})$) denote the total number of intruders captured by the turret that uses an online algorithm $\mathcal{A}$ (resp. optimal offline algorithm $\mathcal{O}$) on an input sequence $\mathcal{I}$. Then, we define the competitive ratio, for an online algorithm, as the following.

\begin{definition}[Competitive Ratio]\label{def:comp_ratio}
Given a problem instance $\mathcal{P}$, an input sequence $\mathcal{I}$, and an online algorithm $\mathcal{A}$, the competitive ratio of $\mathcal{A}$ on $\mathcal{I}$ is defined as $C_{\mathcal{A}}(\mathcal{I},\mathcal{P}) = \tfrac{m_{\mathcal{O}}(\mathcal{I},\mathcal{P})}{m_{\mathcal{A}}(\mathcal{I},\mathcal{P})} \ge 1$, and the competitive ratio of $\mathcal{A}$ for the problem instance $\mathcal{P}$ is $c_{\mathcal{A}}(\mathcal{P}) = \sup_{\mathcal{I}}~ C_{\mathcal{A}}(\mathcal{I},\mathcal{P})$. Finally, the competitive ratio for the problem instance $\mathcal{P}$ is $c(\mathcal{P}) = \inf_{\mathcal{A}} c_{\mathcal{A}}(\mathcal{P})$.
An algorithm is $c$-competitive for the problem instance $\mathcal{P}$ if $c_{\mathcal{A}}(\mathcal{P}) \leq c$, where $c\geq 1$ is a constant.
\end{definition}

Competitive analysis can be viewed as a two-person zero-sum game\footnote{ Determining an optimal competitive ratio is equivalent to determining the value of the game.} between an online player and an adversary \cite{borodin2005online}. The online player operates an online algorithm $\mathcal{A}$ on an input sequence created by the adversary. Conversely, the adversary, with the information of $\mathcal{A}$, constructs an input sequence such that it minimizes the number of intruders captured by $\mathcal{A}$ and simultaneously maximizes the number of intruders captured by $\mathcal{O}$. Thus, we restrict the choice of inputs $\mathcal{I}$ to those for which there exists an optimal offline algorithm $\mathcal{O}$ such that $m_{\mathcal{O}}(\mathcal{I},\mathcal{P})\geq 1,$ over those $\mathcal{I}$. If for some such $\mathcal{I}$, $m_{\mathcal{A}}(\mathcal{I},\mathcal{P}) = 0$, then we say that $\mathcal{A}$ is not $c$-competitive for any finite $c$. Note that the optimal offline algorithm $\mathcal{O}$ designed by the adversary and thus, may be different than the algorithms proposed for problem \textbf{P1}. We now define our second problem.

\textbf{Problem Statement II (P2):} \textit{Design online algorithms for the turret that have finite competitive ratios and establish fundamental guarantees on the existence of online algorithms with finite competitive ratio.}

We start with analyzing problem \textbf{P1} in the next section.

\section{Polynomial Time Control Algorithms for \textbf{P1}}\label{sec:P1}

Consider a linear environment $\mathcal{L}(\theta)$ which is a line segment from $-\theta$ to $\theta$. We first observe that the conical environment $\mathcal{E}(\theta)$ with a turret can be mapped onto a linear environment $\mathcal{L}(\theta)$ of length $2\theta$ with a mobile vehicle, modeled as a point mass, that moves with linear speed $\omega$ in $\mathcal{L}(\theta)$. Thus, our first result establishes that problem \textbf{P1} is equivalent to solving the TRP-TW on a line which is defined as follows.

Consider a line segment from $-L$ to $L$ with a repairperson who seeks to provide service to $N$ locations on the line segment. Each location $i$ requires a service time of $\Delta$ and has a time window $T_i=[s_i,t_i]$ in which it must be serviced, where $s_i$ (resp., $t_i)$ denotes the first (resp. last) time instant at which $i$ is available to the repairperson. The repairperson is allowed to wait at the location of $i$ if it reaches before $s_i$ and must finish servicing $i$ before time $t_i$. The repairperson obtains a unit reward, associated with the service $i$, upon successful completion of service. Then, the TRP-TW problem is to determine a tour through these locations that maximizes the total reward collected by the repairperson.

\begin{theorem}
Problem \textbf{P1} is equivalent to solving the TRP-TW on the line segment $\mathcal{L}(\theta)$.
\end{theorem}
\begin{proof}
The aim is to map the locations and the time intervals for the $N$ intruders in $\mathcal{E}(\theta)$ to the locations in $\mathcal{L}(\theta)$ with each intruder having a particular time window. 
\begin{figure}[t]
     \centering
     \begin{subfigure}[b]{0.45\columnwidth}
         \centering
         \includegraphics[width=\linewidth]{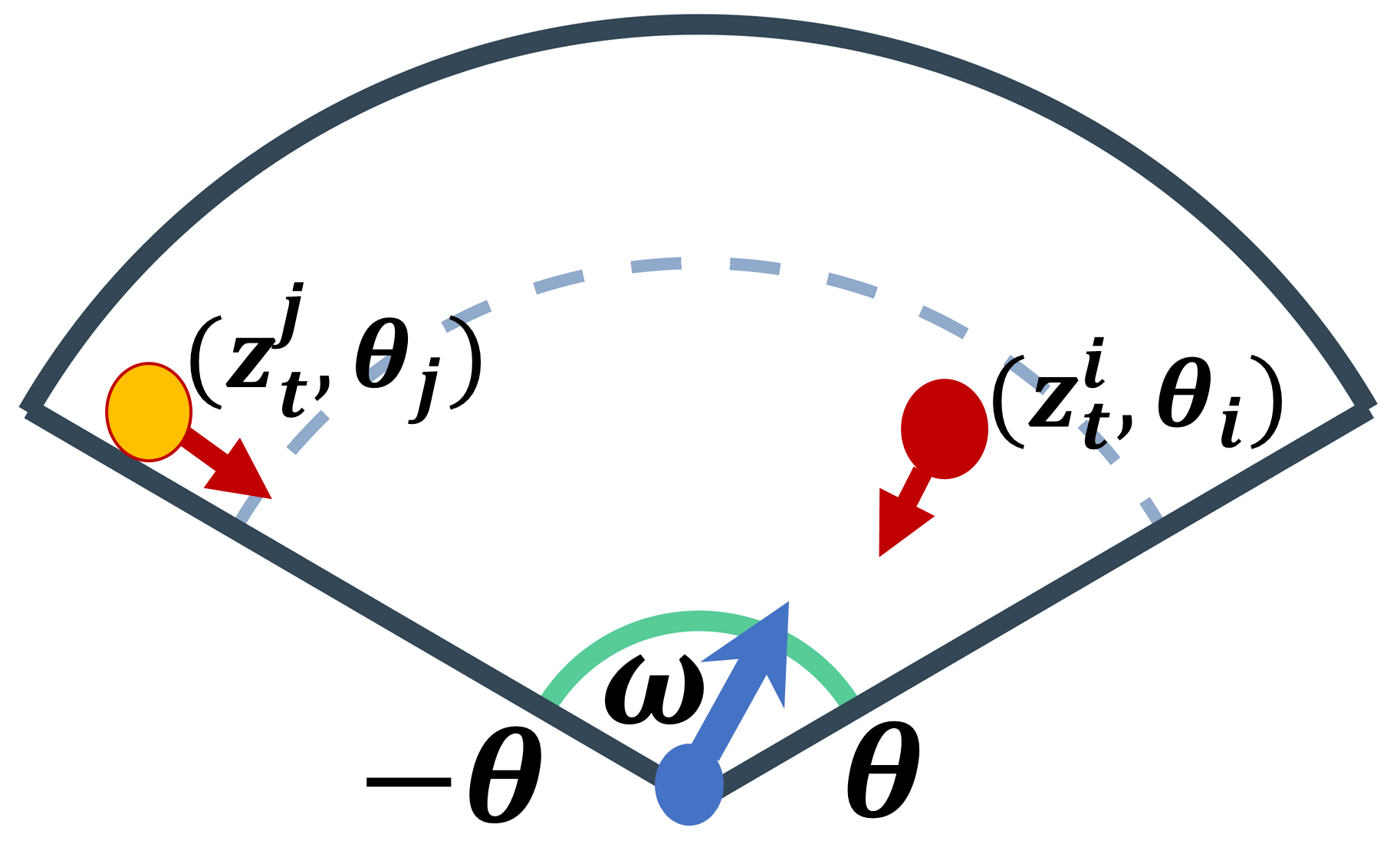}
         \caption{\textbf{P1} in $\mathcal{E}(\theta)$}
         \label{fig:equi_1}
     \end{subfigure}
     \hfill
     \begin{subfigure}[b]{0.45\columnwidth}
         \centering
         \includegraphics[width=\linewidth]{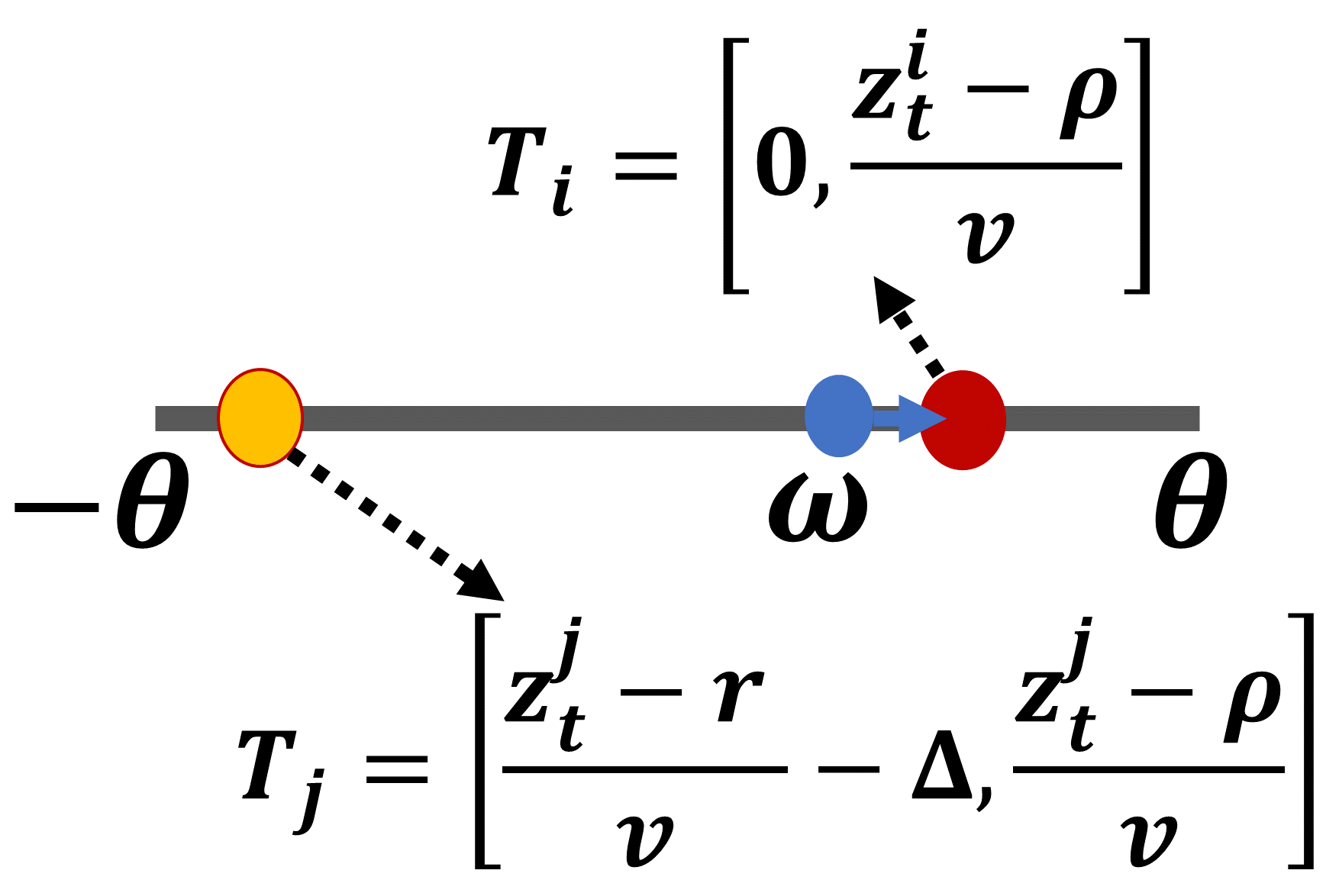}
         \caption{TRP-TW in $\mathcal{L}(\theta)$}
         \label{fig:equi_2}
     \end{subfigure}
     \caption{\small Equivalence between problem \textbf{P1} and TRP-TW. The intruder shown in yellow (resp. red) is outside (resp. within) the range of the turret.}
        \label{fig:equivalence}
        \vspace{-0.2in}
\end{figure}
Since the intruders move radially towards the perimeter, their angular coordinates do not change with time. Therefore, each intruder $i$ located at $(z_t^i,\theta_i)$ in $\mathcal{E}(\theta)$ is mapped to location $\theta_i$ in $\mathcal{L}(\theta)$.

We now show that the total time an intruder $i$ takes to reach the perimeter can be mapped to a time window corresponding to the intruder $i$. Suppose that intruder $i$ is located beyond the radial distance $r+\Delta v$ in $\mathcal{E}(\theta)$. Although the turret can only begin capturing intruder $i$ once it is at most $r+\Delta v$ distance from the origin, the turret does have the information of the angular coordinate of intruder $i$ at time $0$. This implies that the turret has the knowledge of when the intruder $i$ would be within the range.
% Let there be $N_o$ intruders that are located beyond the radial distance $r+\Delta v$ in $\mathcal{E}(\theta)$. Although the turret can only begin capturing any intruder out of the $N_o$ intruders once it is at most $r+\Delta v$ distance from the origin, the turret does have the information of the angular coordinate of the $N_o$ intruders at time $0$. This implies that the turret has the knowledge of when these $N_o$ intruders would be located once they are within the range. 
More formally, an intruder $i$ located at $(z_t^i,\theta_i)$ will be within $r+\Delta v$ distance in time $(\frac{z_t^i-r}{v}-\Delta)$. Further, the same intruder $i$ takes exactly $\frac{z_t^i-\rho}{v}$ time to reach the perimeter (Fig. \ref{fig:equi_1}). Therefore, the same intruder $i$ gets mapped as a static service in $\mathcal{L}(\theta)$ with time window $T_i=[\frac{z_t^i-r}{v}-\Delta,\frac{z_t^i-\rho}{v}]$ and unit reward (Fig. \ref{fig:equi_2}). Note that collocated intruders can be represented as separate individual requests, each with the same time window as the other intruders at the same location and unit reward.
% equal to the number of intruders collocated at location $(z_t^i,\theta_i)$ (Fig. \ref{fig:equi_2}).
% \avm{[
% The reward may not necessarily equal the number of intruders collocated at that location because the turret may not capture all of the intruders since the time required to do so could exceed the time window.
% Earlier, it is mentioned that capturing $n$ collocated intruders requires $n\Delta$ time.
% ]}

We now consider an intruder $i$ with radial distance of at most $r+\Delta v$ from the origin. As the $i$th intruder is already in the range, it is mapped to $\mathcal{L}(\theta)$ with time window defined as $[0,\frac{z_t^i-\rho}{v}]$. This concludes the proof. \qed
\end{proof}

The next result characterizes the complexity of problem \textbf{P1}, proof of which can be found in \cite{nagamochi2008}.

\begin{proposition}
There exists a $2$-approximate algorithm for problem \textbf{P1}.
\end{proposition}
\begin{proof}
An offline TRP-TW problem is considered in \cite{nagamochi2008}, with a difference that the repair person can keep on servicing the $i$th demand after its time window, if the repair person reaches the demand before $t_i$. Thus, the $2$-approximate algorithm in \cite{nagamochi2008} can be applied to this setup with slight modifications.\qed
\end{proof}

\begin{remark}[Special cases]
\begin{itemize}
    \item If $r = \rho$ and $\Delta=0$, then Problem \textbf{P1} is equivalent to the setup from \cite{smith2009dynamic}.
    \item If $\Delta=0$, then Problem \textbf{P1} is equivalent to the setup considered in \cite{bar2005TSPTW}.
\end{itemize} 
\end{remark}
The rest of this section focuses on the special parameter regime of $r = \rho$ and $\Delta > 0$ in which we will show that problem \textbf{P1} admits a polynomial time control algorithm. We begin with the notion of \emph{reachability} of any intruder located at an arbitrary location in the environment. 
% For ease of understanding, we assume that there is one intruder at a particular location. However, the analysis easily extends to multiple collocated intruders. 

% \subsection{Reachability of intruders}
Consider that $r=\rho$. Then, given the orientation of the turret at an angle $\gamma_t$, we say that intruder $i$ located at $(z^i_t,\theta_i)$ is \emph{reachable from the turret} if the time taken by the turret to capture the $i$th intruder does not exceed the time taken by the intruder to reach the perimeter. Mathematically, 
\[
\frac{|\gamma_t - \theta_i|}{\omega} \leq \frac{z^i_t - \rho - \Delta v}{v} \Leftrightarrow z^i_t \geq \rho + \Delta v +\frac{v}{\omega}|\gamma-\theta_i|. 
\]
We generalize this notion to an intruder $j$ being reachable from intruder $i$ if the turret is initially oriented toward $i$, i.e., $\gamma_t = \theta_i$ and needs to capture $j$ after completing the capture of $i$. Mathematically, for $z_t^i\geq \rho+\Delta v$, this is equivalent to
\begin{equation}\label{eq:reachable}
\begin{split}
    &\Delta+\frac{|\theta_j - \theta_i|}{\omega}+\max\{\frac{z_t^i-\rho-\Delta v}{v},0\} \leq \\
    &\frac{z^j_t - \rho - \Delta v}{v} \Leftrightarrow z^j_t-z_t^i \geq v\Delta +\frac{v}{\omega}|\theta_j-\theta_i|.
    % &z_t^i \in [\rho, r+\Delta v] \text{ and } \\
    % &\underbrace{2\Delta}_{\text{Capture $i$ and $j$}}+\underbrace{\frac{|\theta_j - \theta_i|}{\omega}}_{\text{Move to $j$}} \leq \underbrace{\frac{z^j_t - \rho}{v}}_{\text{Time for which $j$ is active}} \\ \Leftrightarrow &z^j_t \geq \rho + 2v\Delta +\frac{v}{\omega}|\theta_j-\theta_i|.
\end{split}
 \end{equation}
% However, if $z_t^i > r+\Delta v$, then there is an additional waiting time of $\frac{z_t^i-r-\Delta v}{v}$ until the $i$th intruder is within the range. Thus, this condition can be compactly written as
% \begin{equation}\label{eq:reachable}
% z^j_t \geq \max\{\frac{z_t^i-r-\Delta v}{v}, 0\} + \rho + 2v\Delta +\frac{v}{\omega}|\theta_j-\theta_i|
% \end{equation}

Now, given a set $I_0$ of initial locations of $N$ intruders, we define a \emph{reachability graph} by representing each intruder as a vertex and creating a directed edge between every pair of vertices $i$ and $j$ if $j$ is reachable from $i$. Since  $v\Delta+\frac{v}{\omega}|\theta_i-\theta_j|$ does not depend on time, the reachability graph does not change with time.
Then, the following result holds and leads to the main result of the section.

% \subsection{Complexity analysis}
% We will first argue that the offline problem is NP-hard for a wide class of problem parameters.

% \begin{theorem}
% The offline problem is NP hard for an arbitrary set of problem parameters.
% \end{theorem}
% \begin{proof}
% The proof of this result relies on two observations. The first is that one can always design a set of initial locations of the intruders so that the reachability graph admits cycles. Second, by selecting these initial locations such that they are all within the range of the turret, we recover the traveling salesperson problem with time windows and non-zero service times, which is known to be NP-hard [REF].  
% \end{proof}

\begin{lemma}\label{lem:special_case}
If $r = \rho$, then both intruder $i$ and $j$ cannot be reachable from each other. 
% \avm{
% Actually, this setting of $r$ contradicts the definition of service time given earlier in the paper ($\Delta \leq \tfrac{r - \rho}{v}$).
% For this setting of $r$ the turret can capture no intruders!
% My guess is that the lemma is supposed to say $r \leq \rho + 2\Delta v$.
% }
\end{lemma}
% \ektmargin{[For this condition, is this equivalent to $r = \rho + \Delta v$ as if $r < \rho + \Delta v$, no intruders can be captured, right? Can we strengthen this condition to $\rho + \Delta v \le r < \rho + 2 \Delta v$?}
\begin{proof}
% As $r\leq \rho+\Delta v$, it follows that both intruder $i$ and $j$ must be located at least $r$ distance away from the origin, i.e., $z_i^t\geq r$ and $z_j^t\geq r$.
From the fact that reachability is defined only when an intruder is captured at a distance $r=\rho$ from the origin, it follows from equation \ref{eq:reachable}, that for any $z_t^j\geq z_t^i$, intruder $i$ is not reachable from $j$ as the turret does not have sufficient time to capture intruder intruder $i$ as $j$ is captured exactly at $\rho$ from the origin.\qed
\end{proof}

% We now present the main result of this section.
\begin{theorem}\label{thm:opt_off}
If $r = \rho$, then a control algorithm that maximizes the number of intruders intercepted is obtained by computing the longest path on the reachability graph. 
\end{theorem}
\begin{proof}
Lemma~\ref{lem:special_case} states that there can be no cycles in the reachability graph of length 2, (i.e., from $i$ to $j$ and back). Further, if $j$ is reachable from $i$, then from equation \eqref{eq:reachable}, we must have $z^j_t > z_t^i$. Therefore, the reachability graph is directed and acyclic. Further, from the definition of reachability, it follows that the reachability graph is time independent. Thus, the optimal number of intruders intercepted is obtained by computing the longest path on this reachability graph, which admits a polynomial time solution using, e.g., the Bellman-Ford algorithm \cite{goldberg1993heuristic}. \qed
\end{proof}

Although Theorem \ref{thm:opt_off} provides an optimal algorithm, the effective parameter regime is limited. Thus, the next section considers an online setup of this problem and focuses on designing online algorithms for worst-case scenarios.

\section{Fundamental Limit and Algorithms For \textbf{P2}}\label{sec:P2}
We first establish a necessary condition on the existence of at best $(N-1)$-competitive algorithms and then design and analyze of online algorithms.
\subsection{Fundamental Limits}

\begin{theorem}\label{thm:N-1-comp}
For any problem instance $\mathcal{P}$ such that $(N-2)(1-\rho)-2(r-\rho)<\frac{2\theta(r-\rho)}{\Delta \omega}$
% \begin{align*}
%     (N-2)(1-\rho)-2(r-\rho)<\frac{2\theta(r-\rho)}{\Delta \omega}
%     \end{align*}
holds, no algorithm can capture all intruders and $c(\mathcal{P})=N-1$
for all choice of intruder speed in
\begin{align*} 
v\in\Big(\frac{\omega(1-\rho)}{2\Delta \omega+2\theta},\frac{r-\rho}{(N-2)\Delta}\Big].
\end{align*}
\end{theorem}
\begin{proof}
From Definition \ref{def:comp_ratio}, the idea is to construct an input sequence for which any online algorithm $\mathcal{A}$ captures at best a single intruder while an optimal offline algorithm $\mathcal{O}$ captures the maximum number of intruders in the same input sequence. We assume that the turret initially starts at angle $\gamma_0 = 0$ for both online and optimal offline algorithms.

The input sequence starts at time instant $\frac{\theta}{\omega}+\Delta$ with a \emph{stream} of intruders, i.e., a single intruder being released every $\max\{\frac{1-\rho}{v}+2\Delta,\frac{2\theta}{\omega}+2\Delta\}$ time units apart, at location $(1,\theta)$. 
If $\mathcal{A}$ never captures any stream intruders, the stream never ends. Thus, algorithm $\mathcal{A}$ does not capture any intruder out of the total $N$ intruders and is therefore, not $c$-competitive for any constant $c\geq 1$. The result then follows as $\mathcal{O}$ can turn the turret, starting at time 0, to $\gamma_{\theta}=\theta$ and capture all the stream intruders.
We thus assume that $\mathcal{A}$ does capture at least one stream intruder, say the $i$th one. Let $t$ denote the time instant when $\mathcal{A}$'s turret captures the $i$th intruder. Note that this means that the turret must remain stationary at angle $\theta$ in the time interval $[t-\Delta,t]$. Then, the input instance ends with the release of a burst of $N-i$ intruders that arrive at location $(1,-\theta)$ at the same time instant $t-\Delta$. For $\theta=\pi$, the burst is released at location $(1,0)$ instead of $(1,-\theta)$.

We now identify how many intruders $\mathcal{A}$ can capture. First, it cannot capture stream intruders 1 through $i-1$ because the stream intruders arrive $\max\{\frac{1-\rho}{v}+2\Delta,\frac{2\theta}{\omega}+2\Delta\}$ time units apart meaning the previous intruder reaches the perimeter and thus is lost before the next stream intruder arrives.
We now show that $\mathcal{A}$'s turret cannot capture any of the $N-i$ burst intruders.
To capture the $i$th intruder, the turret must spend at least $\Delta$ time units at an angle $\theta$. Next, the turret takes exactly $\tfrac{2\theta}{\omega}$ time to move to location $-\theta$ and must spend at least $\Delta$ time units to capture one out of the $N-i$ intruders. Given that $\frac{1-\rho}{v}<2\Delta+\frac{2\theta}{\omega}$ or equivalently $v>\frac{\omega(1-\rho)}{2\Delta\omega+2\theta}$ holds, the turret is ensured to lose the burst intruders. Thus, the input sequence constructed in this proof ensures that any online algorithm captures at best a single intruder out of the $N$ intruders that are released in the environment.

We now determine the number of intruders captured by the optimal offline algorithm
$\mathcal{O}$. 

\textbf{Case 1 ($i=1$):} In this case, $\mathcal{O}$ turns the turret towards $-\theta$ at time $0$. Since the first stream intruder arrives at time instant $\frac{\theta}{\omega}+\Delta$, it is ensured that the turret will be at an angle $-\theta$ at the time instant the burst arrives. More precisely, the burst arrives at least $\Delta$ time units after the turret is at angle $-\theta$. This implies that the turret can start capturing the burst intruders as soon as they are $r+\Delta v$ distance away from the origin. Note than when $r+\Delta v>1$, the turret can still start capturing as $\mathcal{O}$ has the information of the time instant the burst arrives in the environment. Finally, to ensure that the turret captures all $N-1$ burst intruders we require $(N-2)\Delta\leq \frac{r-\rho}{v}$ which yields $v\leq \frac{r-\rho}{(N-2)\Delta}$
% \begin{align*}
%     (N-2)\Delta\leq \frac{r-\rho}{v}
%         \Rightarrow v\leq 
%              \frac{r-\rho}{(N-2
%              )\Delta}.
%     \end{align*}

\textbf{Case 2 ($i>1$):} In this case, $\mathcal{O}$ can turn the turret to angular location $\theta$ until the first $i-1$ intruders have been captured and then move the turret to location $-\theta$ to capture the burst of $N-i$ intruders. The explanation can be summarized as follows.
\begin{itemize}
    % \item In order to reach angular location $-\theta$ in the least time, the turret following $\mathcal{O}$ must capture $1,\dots,i-1$ intruders as soon as they are within the range, i.e., at a radial distance $r$.

    \item Assuming that the online algorithm captures the $i$th intruder at a distance $r$ from the origin, the additional time that the optimal offline algorithm gets before the burst arrives is exactly $\max\{\tfrac{1-\rho}{v}+\Delta,\tfrac{2\theta}{\omega}+\Delta\}$. Similar to Case 1, this implies that turret will have captured one intruder out of the burst by the time the burst is $r$ distance away from the origin.
    
    \item Finally, to ensure that the optimal offline algorithm can capture all of the $N-i$ intruders in the burst, we require $(N-i-1)\Delta\leq \frac{r-\rho}{v}$ as each intruder requires $\Delta$ amount of service time.
    % \begin{align*}
    % (N-i-1)\Delta\leq \frac{r-\rho}{v}
    % \Rightarrow v\leq \frac{r-\rho}{(N-i-1)\Delta}.
    % \end{align*}
\end{itemize}
Note that $i=1$ yields the least value of $v$.
Thus, we have established that any online algorithm cannot capture more than one intruder and the optimal offline algorithm captures all $N-1$ intruders for any $v\in (\frac{\omega(1-\rho)}{2\Delta \omega+2\theta},\frac{r-\rho}{(N-2)\Delta}]$. Finally the range of $v$ is well defined only if $\frac{\omega(1-\rho)}{2\Delta \omega+2\theta}<\frac{r-\rho}{(N-2)\Delta}$. This concludes the proof. \qed
\end{proof}

\begin{remark}
A family of upper bounds on $v$ in Theorem \ref{thm:N-1-comp} can be obtained by relaxing the requirement that $\mathcal{O}$ must capture all burst intruders in Case 1. Specifically, for every $1<j<N$, the relaxed requirement $(N-j-1)\Delta\leq \frac{r-\rho}{v}$ yields $c(\mathcal{P})=N-j$.
\end{remark}

\subsection{Online Algorithms}
In this section, we present two online algorithms with finite competitive ratios in certain parameter space. 

We define an epoch as the time interval that starts at the time instant the turret turns, either clockwise or anti-clockwise, from its starting location and ends when the turret is just about to start its next epoch upon returning to its starting location. We denote the start of an epoch $k$ by $k_s$.

\subsubsection{\textbf{Sweeping Turret (SiT)}}
Algorithm SiT is an open loop and memoryless algorithm having the best competitive ratio, i.e., $c_{SiT}=1$, and is summarized for $\theta<\pi$ as follows. 

At time instant $0$ or equivalently in the first epoch, the turret starts at angle $\gamma_0=-\theta$ and turns toward angle $\theta$ at every time instant. If, at time instant $t$, there exists $1\leq n\leq N$ intruders collocated at location $(z_t^n,\theta_n)$ such that $z_t^n\leq \min\{v\Delta +r,1\}$ and $\phi_n=\gamma_t$, then the turret locks onto the $n$ intruders at time $t$. The turret captures the $n$ intruders until time $t+\Delta n$ and then continues to turn towards $\theta$. Upon reaching angle $\theta$, the turret changes its direction and moves towards $-\theta$ capturing intruders that are at most $\Delta v+r$ distance from the turret and have the same angle. The next epoch begins after the turret reaches $-\theta$.

If $\theta=\pi$, then the turret does not change its direction upon reaching $\theta$. Instead, it keeps turning at all time instances in a circular path. 

We now establish the parameter regime under which Algorithm SiT is $1$-competitive.
\begin{theorem}
For any problem instance $\mathcal{P}$, $c_{SiT}=1$ if 
\begin{align*}
    &v\leq \min\Big\{\frac{1-r}{\Delta},\frac{\omega(r-\rho)}{a\theta+(N-1)\Delta\omega}\Big\} \text{ or } \\
    &\frac{1-r}{\Delta}<v\leq \frac{\omega(1-\rho)}{a\theta+N\Delta \omega},
\end{align*}
where $a=4$ when $\theta<\pi$ and $a=2$, otherwise.
% For problem instance $\mathcal{P}$ such that 
% \begin{align*}
%     v\leq
%     \begin{cases}
%     \frac{\omega(r-\rho)}{a\theta+(N-1)\Delta\omega}, \text{ if } r+\Delta v<1\\
%     \frac{\omega(1-\rho)}{a\theta+N\Delta \omega} \text{ otherwise. }
%     \end{cases}
% \end{align*} holds, Algorithm SiT is $1$-competitive, where $a=4$ when $\theta<\pi$ and $a=2$, otherwise. 
\end{theorem}
\begin{proof}
Without loss of generality, assume that the turret has just left angle $\gamma_t=-\theta$ at some time instant $t$ and we consider that $\theta<\pi$. The proof for $\theta=\pi$ is analogous and has been omitted for brevity. To construct the worst case, let there be $n_1>0$ collocated intruders at time instant $t-1$, all at angle $-\theta$ and at a radial distance of $\min\{r+\Delta v+\epsilon,1\}$, where $\epsilon>0$ is a very small number. The time taken by the turret to turn to angle $\theta$ and then back to $-\theta$ is exactly $\frac{4\theta}{\omega}+n_2\Delta$, where $0\leq n_2<N-n_1$ denotes the number of intruders that the turret captures along its path. For Algorithm SiT to be $1$-competitive, we require that no intruder must be lost. Mathematically, we require 
\begin{align*}
    \frac{\min\{1,r+\Delta v\}-\rho}{v}\geq \frac{4\theta}{\omega}+(n_1+n_2)\Delta
\end{align*}
where we used the fact that the turret requires $n_1\Delta$ amount of time to capture the $n_1$ intruders upon returning to angle $-\theta$. In the worst case, $n_1+n_2=N$ holds which yields $\frac{\min\{r+\Delta v,1\}-\rho}{v}\geq \frac{4\theta}{\omega}+N\Delta$
% \begin{align*}
%     &\frac{\min\{r+\Delta v,1\}-\rho}{v}\geq \frac{4\theta}{\omega}+N\Delta
    % &\Rightarrow v\leq
    % \begin{cases}
    % \frac{\omega(r-\rho)}{4\theta+(N-1)\Delta\omega}, \text{ if } r+\Delta v<1\\
    % \frac{\omega(1-\rho)}{4\theta+N\Delta \omega} \text{ otherwise. }
    % \end{cases}
% \end{align*}
and the proof is complete. \qed
\end{proof}

Although Algorithm SiT is $1$-competitive, it does not consider the intruders with radial coordinate beyond $r+\Delta v$ to plan its motion. This yields a conservative parameter regime for which Algorithm SiT is effective. This motivates our next algorithm which is memoryless, but not open loop.

\begin{figure}[t]
    \centering
    \includegraphics[scale=0.21]{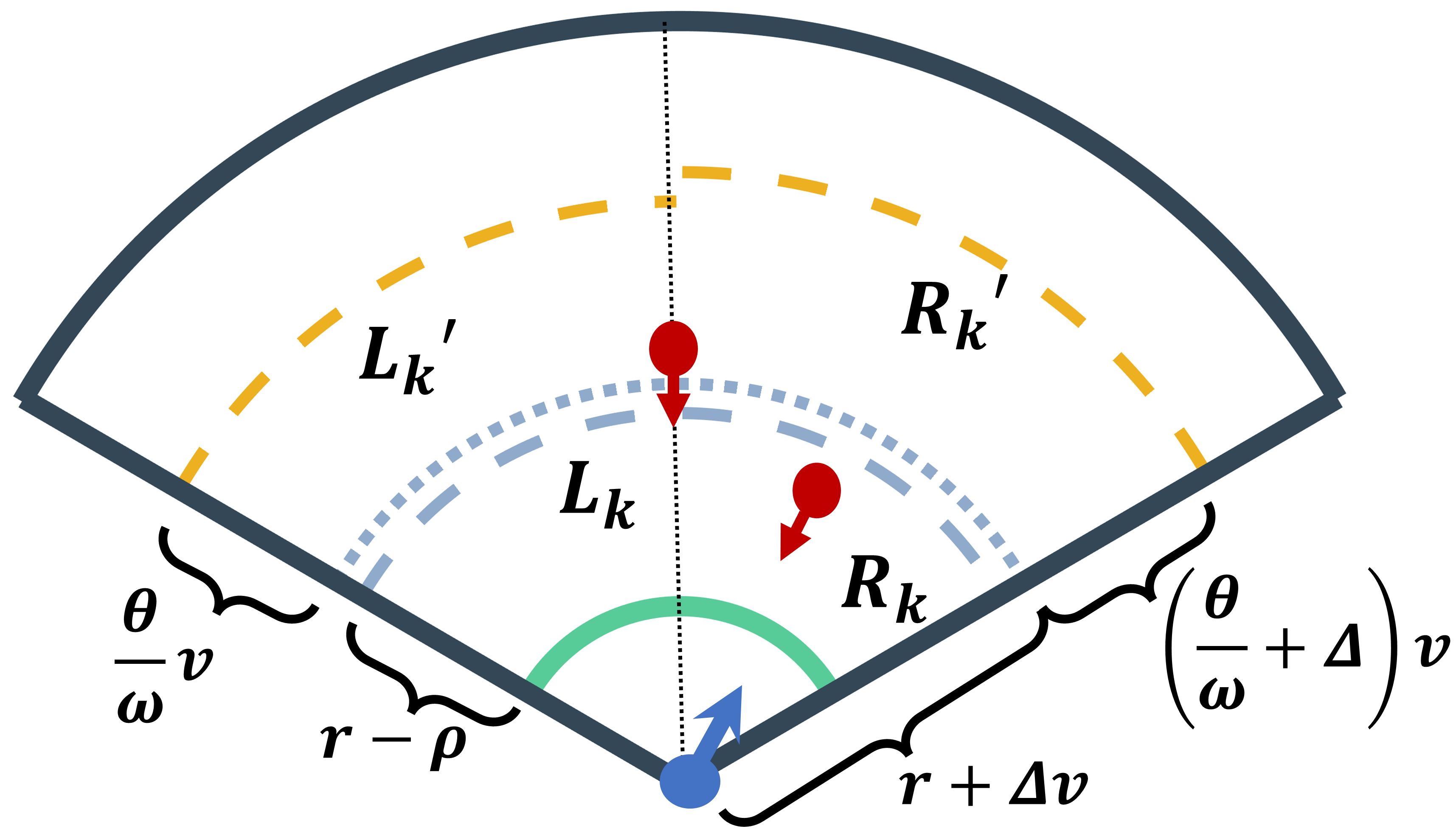}
    \caption{\small Description of DPaC Algorithm. The blue dashed curve denotes the range of the turret. The region between the blue dot curve and the yellow dashed curves denotes the set $R_k'$ and $L_k'$. The region between the perimeter (green curve) and the blue dot curve denotes the set $R_k$ and $L_k$. Note that the yellow dashed curve of $R_k'$ is higher than that of $L_k'$ because $|R_k'|>|L_k'|$.}
    \label{fig:DPaC}
        \vspace{-0.2in}
\end{figure}

\subsubsection{\textbf{Dynamically Project and Capture (DPaC)}}

The intuition behind this algorithm is to partition the environment into two halves and capture intruders from the side which has higher number of intruders. 
However, the time taken by the turret to return to its starting location is a function of the number of intruders it captures on its way and thus, is not known at time $0$. Thus, at the start of every epoch, the turret selects the side by determining the time taken by the turret to reach either $\theta$ or $-\theta$. This is further explained below.

To compare the number of intruders on either side of the turret, we define four sets of intruders, $R_k,L_k,R_k'$, and $L_k'$ for an epoch $k$ and we denote $|S|$ as the cardinality of set $S$. Set $R_k$ (resp. $L_k$) is the set of intruders that are at most $r+\Delta v$ distance at time instant $k_s$ and with angular coordinates in $[0,\theta]$ (resp. $[-\theta,0)$) (Fig. \ref{fig:DPaC}). 
% Set $R_k'$ (resp $L_k'$) is the set of intruders that will be in the range once the turret reaches $\theta$ (resp. $-\theta$). More precisely, s
Similarly, set $R_k'$ (resp. $L_k'$) is the set of intruders with radial coordinate strictly more that $r+\Delta v$ and at most $\min\{1,r+(\frac{\theta}{\omega}+(|R_k|+1)\Delta) v\}$ (resp. $\min\{1,r+(\frac{\theta}{\omega}+(|L_k|+1)\Delta )v\}$) and angular coordinate in $[0,\theta]$ (resp. $[-\theta,0)$). Note that the sets $R_k'$ and $L_k'$ require that $r+\Delta v<1$. If $r+\Delta v\geq 1$, then $R_k'=L_k'=\emptyset$. 
% In this case, the turret returns to its starting location without capturing any intruder while turning from $\theta$ or $-\theta$ to $0$.

We now summarize Algorithm DPaC which is formally defined in Algorithm \ref{algo:DPaC}. The turret starts at angle $0$ at the start of every epoch $k$. Algorithm DPaC compares the total number of intruders at either side of the turret at time instant $k_s$ and contained in the four sets. If $|R_k|+|R_k'|\geq |L_k|+|L_k'|$ holds at time $k_s$, then the turret turns clockwise towards angle $\theta$. Otherwise, the turret turns anti-clockwise towards angle $-\theta$. While turning towards $\theta$ (resp. $-\theta$), the turret captures only the intruders that are in $R_k$ (resp. $L_k$). After reaching  $\theta$ (resp. $-\theta$), the turret turns toward the starting location capturing intruders that were contained in $R_k'$ (resp. $L_k'$) at time $k_s$. Epoch $k+1$ begins once the turret reaches angle $0$. We assume that all of the intruders that the turret decided not to capture in epoch $k$ are lost by time $(k+1)_s$.

\begin{algorithm}[t]
\DontPrintSemicolon
	\SetAlgoLined
	\For{each epoch $k\geq1$}{
	Determine sets $R_k,R_k',L_k,L_k'$\\
	\eIf{$|R_k|+|R_k'|\geq|L_k|+|L_k'|$}{
	    Turn towards $\theta$ and capture intruders in $R_k$\\
	   Turn towards $0$ and capture intruders in $R_k'$
		\;}
	{ 
	Turn towards $-\theta$ and capture intruders in $L_k$\\
	Turn towards $0$ and capture intruders in $L_k'$
	\;
	}}
	\caption{Dynamically Project and Capture}
	\label{algo:DPaC}
\end{algorithm}

We now characterize the parameter regime of Algorithm DPaC. For ease of understanding, we only focus on input sequences that have equal number of intruders arriving on both sides of the turret in every epoch. If $N$ is odd, then that means one side has one intruder more than the other side in at most one epoch. Later, we will show that the bound holds for any input sequence.

\begin{lemma}\label{lem:DPaC_cond}
Given a problem instance $\mathcal{P}$, suppose that
\begin{align*}
    &v\leq \min\Big\{\frac{1-r}{\Delta},\frac{\omega(1-\rho)}{3\theta+\lceil0.5N\rceil\Delta\omega}, \frac{\omega(r-\rho)}{2\theta+(\lceil0.5N\rceil-1)\Delta\omega}\Big\}\\
    &\text{ or } \frac{1-r}{\Delta}< v\leq \frac{\omega(1-\rho)}{3\theta+\lceil0.5N\rceil\Delta\omega}.
\end{align*}
Further, suppose that the number of intruders is equal on each side in every epoch. Then, any intruder that is not considered for comparison in epoch $k$ is 
\begin{enumerate}[(i)]
    \item considered for comparison in epoch $k+1$ and
    \item is capturable in epoch $k+1$.
\end{enumerate}

% \begin{align*}
%     v\leq \begin{cases}
%   \min\{\frac{\omega(1-\rho)}{3\theta+\lceil0.5N\rceil\Delta\omega}, \frac{\omega(r-\rho)}{2\theta+(\lceil0.5N\rceil-1)\Delta\omega}\} \text{ if } r+\Delta v < 1\\
%     \frac{\omega(1-\rho)}{3\theta+\lceil0.5N\rceil\Delta\omega}, \text{ otherwise }.
%     \end{cases}
% \end{align*}
\end{lemma}
\begin{proof}
Given the assumption, $|R_k|+|R_k'| = |L_k|+|L_k'|$ holds for an epoch $k$. Two cases arise:

\textbf{Case 1 ($r+\Delta v<1$):} Suppose that $|R_k|\geq |L_k|$. The proof is analogous for $|R_k|< |L_k|$. Let there be $C_l$ intruders that are located at $(\min\{1,r+(\frac{\theta}{\omega}+(|L_k|+1)\Delta )v+\epsilon\},-\theta)$, where $\epsilon$ is a very small positive number. Given the location of these $C_l$ intruders, they will not be considered for comparison in epoch $k$. The time taken by the turret to complete epoch $k$ and return to $\gamma_{(k+1)_s}=0$ is $\frac{2\theta}{\omega}+(|R_k|+|R_k'|)\Delta$. To ensure that these $C_l$ intruders can be captured by the turret in epoch $k+1$, it suffices to show that the $C_l$ intruders do not reach the perimeter by the time the turret captures them. Note that if these $C_l$ intruders are not lost in epoch $k+1$, then that implies that they were contained in either set $L_{k+1}$ or $L_{k+1}'$ at time $(k+1)_s$. The turret requires at most $\frac{\theta}{\omega}+(|L_{k+1}|+C_l)\Delta$ time to capture the $C_l$ intruders in epoch $k+1$. Note that the $C_l$ intruders will be contained in the set $L_{k+1}$ at time $(k+1)_s$ as $\tfrac{r+\tfrac{\theta}{\omega}+(|L_k|+1)\Delta)v-r-\Delta v}{v}\leq \tfrac{r+\tfrac{\theta}{\omega}+(|R_k|+1)\Delta)v-r-\Delta v}{v}$ holds. Thus, they will be considered for comparison in epoch $k+1$.
For successful capture of $C_l$ intruders, the following condition must hold.
\begin{equation}\label{eq:DPaC_cond}
\begin{split}
    &\frac{2\theta}{\omega}+(|R_k|+|R_k'|)\Delta +\frac{\theta}{\omega}+(|L_{k+1}|)\Delta\leq\\
    &\frac{\min\{1,r+(\frac{\theta}{\omega}+(|L_k|+1)\Delta)v\}-\rho}{v}.
    % &\Rightarrow \frac{2\theta}{\omega}+\Delta(|R_k|+|R_k'|+|L_{k+1}|+C_l-|L_k|-1)\leq \\
    % &\frac{r-\rho}{v}
\end{split}
\end{equation}
Note that $|L_k|=0$ in the worst case and $L_{k+1}$ contains $C_l$. Two cases arise.

\textbf{Case 1.1 ($1<r+(\frac{\theta}{\omega}+(|L_k|+1)\Delta)v)$:}
The assumption of equal number of intruders arriving on both sides implies that $|R_k|+|R_k'|=|L_k|+|L_k'|$ must hold in epoch $k$. Similarly, if the turret turns towards $-\theta$ in epoch $k+1$, then $|L_{k+1}|-C_l+|L_{k+1}'|+C_l>|R_{k+1}|+|R_{k+1}'|$ must hold at time $(k+1)_s$. Finally, since the total number of intruders is at most $N$ yields $|R_k|+|R_k'|+|L_k|+|L_k'|+|L_{k+1}|+|L_{k+1}'|+|R_{k+1}|+|R_{k+1}'|\leq N$. Using the fact that same number of intruders arrive in the environment and that $|L_{k+1}'|=0$ in the worst-case, it follows that $|R_k|+|R_k'|+|L_{k+1}|\leq \lceil0.5N\rceil$, where $\lceil.\rceil$ denotes the ceil function and has been used as the number of intruders captured by the turret is an integer. 
Thus, equation \eqref{eq:DPaC_cond} is guaranteed if $v\leq \tfrac{\omega(1-\rho)}{3\theta+\Delta\omega(\lceil0.5N\rceil)}$.
% \begin{align*}
%     v\leq \frac{\omega(1-\rho)}{3\theta+\Delta\omega(\lceil0.5N\rceil)}.
% \end{align*}

\textbf{Case 1.2 ($1\geq r+(\frac{\theta}{\omega}+(|L_k|+1)\Delta)v)$):}
Similar to Case 1.1, using the fact that equal number of intruders arrive on both sides, it follows that $|R_k|+|R_k'|+|L_{k+1}|\leq \lceil 0.5N\rceil$ which yields $v\leq \tfrac{\omega(r-\rho)}{2\theta+\Delta\omega(\lceil0.5N\rceil-1)}$ as a sufficient condition.
% \begin{align*}
%     v\leq \frac{\omega(r-\rho)}{2\theta+\Delta\omega(\lceil0.5N\rceil-1)}.
% \end{align*}

\textbf{ Case 2 ($r+\Delta v\geq 1$):} In this case, the turret returns to its starting location, without capturing any intruders, upon reaching $\theta$ or $-\theta$. Similar to Case 1, suppose that $|R_k|\geq |L_k|$. Let $C_l$ intruders arrive at location $(1,-\theta)$ as soon as the turret starts turning towards angle $\theta$ in epoch $k$. The $C_l$ intruders require exactly $\frac{1-\rho}{v}$ time to reach the perimeter. As $R_k'=L_k'=\emptyset$, for successful capture of $C_l$ intruders and using the fact that equal numbers of intruders arrive on both sides in the worst case yields $\frac{3\theta}{\omega}+(|R_k|+|L_{k+1}|)\Delta \leq \frac{1-\rho}{v}$. This simplifies to the condition $v\leq \frac{(1-\rho)\omega}{3\theta+\lceil0.5N\rceil\Delta\omega}$.
% \begin{align*}
%     % &\frac{2\theta}{\omega}+R_k\Delta+\frac{\theta}{\omega}+(L_{k+1}+C_l)\Delta \leq \frac{1-\rho}{v}\\
%     \frac{3\theta}{\omega}+(|R_k|+|L_{k+1}|+C_l)\Delta \leq \frac{1-\rho}{v}
%     \Rightarrow v\leq \frac{(1-\rho)\omega}{3\theta+\lceil0.5N\rceil\Delta\omega}.
% \end{align*}
Finally, if the $C_l$ intruders are not lost, then this implies that they are considered for comparison at time $(k+1)_s$. This concludes the proof. \qed
\end{proof}

An important consequence of Lemma \ref{lem:DPaC_cond} is that any intruder that is contained in $L_k$ or $R_k$ can be captured in epoch $k$. We now show that this holds even for intruders in the set $R_k'$ and $L_k'$.

\begin{corollary}\label{cor:DPaC}
Suppose that the turret decides to turn towards $\theta$ (resp. $-\theta$) at time instant $k_s$. Then, the turret captures all intruders in the set $R_k'$ (resp. $L_k'$) by the end of epoch $k$ in the parameter regimes of Lemma \ref{lem:DPaC_cond}.
\end{corollary}
\begin{proof}
Without loss of generality, suppose that $|R_k|+|R_k'|\geq |L_k|+|L_k'|$ holds at time $k_s$. We only consider that $r+\Delta v<1$ as $R_k'=L_k'=\emptyset$ for $r+\Delta v\geq 1$. The time taken by the turret to turn to angle $\theta$ from the starting position is exactly $\frac{\theta}{\omega}+\Delta |R_k|$. In the worst case, $0<m\leq R_k'$ out of the total $R_k'$ intruders are located at $(r+\Delta v+\epsilon,0)$, where $\epsilon>0$ is a very small number. Thus, in order to successfully capture all $m$ intruders by the end of epoch $k$, the condition $\frac{2\theta}{\omega}+(|R_k|+|R_k'|)\Delta\leq \frac{r-\rho}{v}+\Delta$ must hold. Since the parameter regime in Lemma \ref{lem:DPaC_cond} holds, the capture of all $|R_k'|$ is guaranteed, and the result is established. \qed
% \begin{align*}
%     &\frac{2\theta}{\omega}+(|R_k|+|R_k'|)\Delta\leq \frac{r-\rho}{v}+\Delta\\
%     &\Rightarrow v\leq \frac{\omega(r-\rho)}{2\theta+(\lceil 0.5N\rceil-1)\Delta\omega}
% \end{align*}
% which holds if Lemma \ref{lem:DPaC_cond} holds. 
\end{proof}

% \begin{corollary}\label{cor:DPaC_2}
% For any problem instance for which Lemma \ref{lem:DPaC_cond} holds, every intruder in $R_k$ (resp. $L_k$) is captured in epoch $k$
% \end{corollary}
% \begin{proof}
% Without loss of generality, suppose that $|R_k|+|R_k'|\geq |L_k|+|L_k'|$ holds at time $k_s$. 

% We first consider that $r+\Delta v\geq 1$. In the worst-case, all $|R_k|$ intruders arrive at location $(1,\theta)$ as soon as the turret leaves angle $0$ in epoch $k-1$. Since $r+\Delta v\geq 1$, the turret does not capture any intruder while turning back to starting location from $\theta$. Thus, the total time required by the turret to capture all $R_k$ intruders in epoch $k$, starting from time instant $(k-1)_s$, is $\frac{3\theta}{\omega}+|R_k|\Delta$ whereas the intruders require $\frac{1-\rho}{v}$ time to reach the perimeter. For successful capture, the condition $v\leq \frac{\omega(1-\rho)}{3\theta+|R_k|\Delta\omega}$ must hold. This condition holds if Lemma \ref{lem:DPaC_cond} holds as $|R_k|\leq \lceil0.5N\rceil$. 
% We now consider $r+\Delta v<1$. Observe that any intruder which is contained in the set $R_k$ was not considered for comparison in epoch $k-1$. Otherwise, these intruders would have been contained in the set $R_{k-1}'$ and had been captured in epoch $k-1$.
% Then, from Lemma \ref{lem:DPaC_cond}, it follows that the turret captures all $|R_k|$ intruders in epoch $k$.
% \end{proof}
We now relax the assumption that equal number of intruders arrive on both sides of the turret.
\begin{corollary}\label{cor:atleast_half}
In the parameter regime specified in Lemma \ref{lem:DPaC_cond}, Algorithm DPaC captures at least $\lceil0.5N\rceil$ intruders in input sequences even with unequal number of intruders arriving on either side.
\end{corollary}
\begin{proof}
Assume that $|R_k|+|R_k'|\geq |L_k|+|L_k'|$ holds for an epoch $k$. As the proof is analogous to the proof of Lemma \ref{lem:DPaC_cond}, we only establish this result for Case 1.1 of Lemma \ref{lem:DPaC_cond}. The explanation is analogous for all other cases and thus has been omitted for brevity. 

Recall that equation \eqref{eq:DPaC_cond}  for Case 1.1 is $\frac{3\theta}{\omega} + (|R_k|+|R_k'|+|L_{k+1}|)\Delta \leq \frac{1-\rho}{v}$. If $|R_k|+|R_k'|+|L_{k+1}|\leq \lceil 0.5N\rceil$ in epoch $k$, then the turret captures all $|R_k|+|R_k'|+|L_{k+1}|$ intruders in the parameter regime of Lemma \ref{lem:DPaC_cond}. Thus, if $|R_k|+|R_k'|+|L_{k+1}|\leq \lceil 0.5N\rceil$ holds for every epoch $k$, then because the turret turns towards the side which contains higher number of intruders in every epoch, it is guaranteed to capture at least $\lceil 0.5 N\rceil$ intruders. For some epoch $k$, if $|R_k|+|R_k'|+|L_{k+1}|>\lceil0.5N\rceil$ holds, in the regime of Lemma \ref{lem:DPaC_cond}, it is ensured that the turret captures at least $\lceil0.5N\rceil$ intruders out of $|R_k|+|R_k'|+|L_{k+1}|$. This concludes the proof.\qed
% Now suppose that $|R_k|+|R_k'|=m_1\geq 0$ and $|L_k|+|L_k'|=m_2$ such that $m_1 > m_2$. This yields $\frac{3\theta}{\omega} + (N-m_2)\Delta \leq \frac{1-\rho}{v}$. As $m_1>m_2$, we have $N-m_2>\lceil0.5N\rceil-m_1$.

% To ensure that Algorithm DPaC to be $2$-competitive, it suffices for the turret to capture $\lceil0.5N\rceil$ intruders under the parameter space characterized by Lemma \ref{lem:DPaC_cond}. Thus, given that the turret captured $m_1$ intruders in epoch $k$, it suffices for the turret to capture $\lceil0.5N\rceil-m_1$ intruders out of the total $N-m_2$ intruders. Thus, the total time required by the turret to capture $\lceil0.5N\rceil-m_1$ intruders, given that it captures $m_1$ intruders in epoch $k$, is $\frac{3\theta}{\omega} + \lceil0.5N\rceil\Delta$. Thus, for successful capture, we require $\frac{3\theta}{\omega} + \lceil0.5N\rceil\Delta\leq \frac{1-\rho}{v}$
% which holds if Lemma \ref{lem:DPaC_cond} holds. This concludes the proof.
\end{proof}

\begin{theorem}\label{thm:DPaC}
For any $\mathcal{P}$, $c_{DPaC}=2$ in the parameter regimes specified in the statement of Lemma \ref{lem:DPaC_cond}.
\end{theorem}
\begin{proof}
In every epoch $k$, Algorithm DPaC turns the turret to the side which has higher number of intruders. Lemma \ref{lem:DPaC_cond} ensures that every intruder is considered for comparison. Further, from Lemma \ref{lem:DPaC_cond} and Corollary \ref{cor:DPaC}, it is ensured that the turret captures either all of the intruders contained in $R_k+R_k'$ or $L_k+L_k'$. Finally, Corollary \ref{cor:atleast_half} ensures that at least $\lceil 0.5N\rceil$ intruders are captured. Thus, by assuming that an optimal offline algorithm exists which captures all $N$ intruders, we obtain the result.\qed
\end{proof}

In the next subsection, we provide a parameter regime plot to highlight the effective parameter space of the algorithms.

\subsection{Numerical Observations}
\begin{figure}[t]
    \centering
    \includegraphics[scale=0.5]{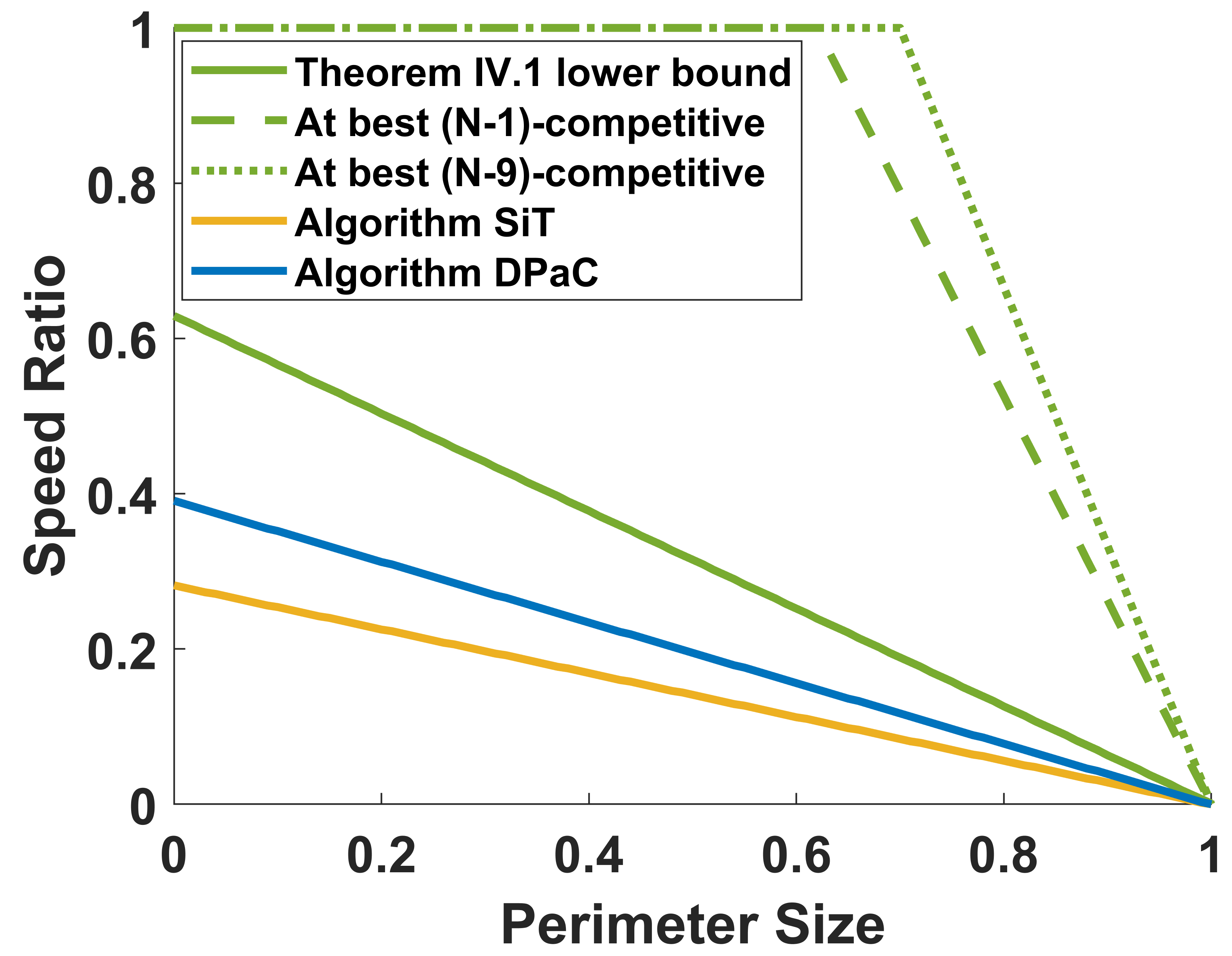}
    \caption{\small Parameter space plot for $N=40, \omega=1,\theta=\frac{\pi}{4},\Delta=0.01$ and $r=1$. Algorithm SiT is $1$-competitive for any value of $v$ below the yellow curve. Algorithm DPaC is $2$-competitive for any value of $v$ below the blue curve. There exists an algorithm with at best $(N-1)$-competitive (resp. $(N-9)$-competitive) ratio between the green solid and the green dashed (resp. dotted) curves. }
    \label{fig:v_rho}
    \vspace{-0.2in}
\end{figure}

We now provide a numerical visualization of the analytic bounds derived in this paper. Figure \ref{fig:v_rho} shows the $(\rho,v)$ parameter regime plot for fixed value of $N,r,\theta,\omega$ and $\Delta$. 

Given the value of parameters in Fig. \ref{fig:v_rho}, there is a very small region in which there could exist an algorithm with competitive ratio better than $N-1$ (below the solid green curve). Recall from proof of Theorem \ref{thm:N-1-comp} that the optimal offline algorithm cannot capture all $N-1$ intruders. Thus, by relaxing the upper bound on $v$ in Theorem \ref{thm:N-1-comp}, there may exist algorithms with competitive ratio better than $N-1$. This is because the number of intruders captured by the optimal offline algorithm decreases in those parameter regimes. For instance, there exists an algorithm which is at best $(N-9)$-competitive between the green dashed and the dotted curve (Fig. \ref{fig:v_rho}). For any value of $v$ which is below the yellow curve, Algorithm SiT is $1$-competitive and for any value of $v$ below the blue curve, Algorithm DPaC is $2$-competitive. For values of $\rho>0.995$, the curve for Algorithm DPaC and Algorithm SiT overlap, meaning that Algorithm SiT is more effective than Algorithm DPaC for $\rho>0.995$.

\section{Conclusion and Future Directions}\label{sec:conc}
This work analyzed a perimeter defense problem in which a single turret, having a finite range and service time, is tasked to defend a perimeter against at most $N$ intruders that arrive in the environment. An offline as well as an online version of this setup was considered. In the offline setup in which $N$ intruders have already arrived in the environment, we established that the problem is equivalent to solving a Travelling Repairperson Problem with Time Windows. We then provided a $2$ approximate algorithm for any value of parameters and a control algorithm that runs in polynomial time for a specific parameter regime. In the online setup, we designed and analyzed two classes of online algorithms and characterized parameter regimes in which they exhibit finite competitive ratios. A necessary condition on the existence of at best $N-1$-competitive algorithms was also established.

Apart from closing the gap between the curve defined by Theorem \ref{thm:DPaC} and Theorem \ref{thm:N-1-comp}, key future directions include multi-vehicle scenarios with energy constraints. Analyzing the problem with a weaker model of the adversary or the turret with a \emph{look-ahead} are also potential extensions.
\bibliographystyle{IEEEtran}
\bibliography{references}
\end{document}